\documentstyle[aps,floats,prd,psfig,amsmath,epsfig]{revtex}

\newcommand{\ApJL}{Astrophys. J. Lett.}
\newcommand{\ApJ}{Astrophys. J.}
\newcommand{\PRL}{Phys. Rev. Lett.}
\newcommand{\PRD}{Phys. Rev. D}
\newcommand{\MNRAS}{Mon. Not. R. Astron. Soc.}
\newcommand{\AsAs}{Astron. Astrophys.}
\newcommand{\AnnAsAs}{Ann. Rev. Astron. Astrophys.}
\newcommand{\aut}[2]{{#2.\ #1}}
\newcommand{\refs}[6]{#2, {\bf #3} {#4} (#5)}

\newcommand{\amp}{and }

\newcommand{\tot}{{\rm t}}

\newcommand{\cmb}{\Theta}

\newcommand{\n}{{\rm n}}
\newcommand{\vecl}{{\bf l}}
\newcommand{\vecla}{{{\bf l}_1}}
\newcommand{\veclb}{{{\bf l}_2}}
\newcommand{\veclc}{{{\bf l}_3}}
\newcommand{\vecld}{{{\bf l}_4}}
\newcommand{\vecka}{{{\bf k}_1}}
\newcommand{\veckb}{{{\bf k}_2}}
\newcommand{\veckc}{{{\bf k}_3}}
\newcommand{\veckd}{{{\bf k}_4}}
\newcommand{\vecke}{{{\bf k}_5}}
\newcommand{\veckf}{{{\bf k}_6}}
\newcommand{\veckg}{{{\bf k}_7}}
\newcommand{\veckh}{{{\bf k}_8}}
\newcommand{\vecki}{{{\bf k}_i}}
\newcommand{\intl}[1]{\int \frac{d^2 {\bf l}_#1}{(2\pi)^2}}
\newcommand{\intk}[1]{\int \frac{d^2 {\bf k}_#1}{(2\pi)^2}}
\newcommand{\intln}{\int \frac{d^2 {\bf l}}{(2\pi)^2}}
\newcommand{\intlnp}{\int \frac{d^2 {\bf l'}}{(2\pi)^2}}
\newcommand{\intlnpp}{\int \frac{d^2 {\bf l''}}{(2\pi)^2}}
\newcommand{\VE}{\frac{d^2 {\bf l}}{(2\pi)^2}}
\newcommand{\bfl}{{\mathbf{l}}}
\newcommand{\bfd}{{\mathbf{d}}}
\newcommand{\bflp}{{\mathbf{l^{\prime}}}}
\newcommand{\bflpp}{{\mathbf{l^{\prime\prime}}}}
\newcommand{\bfL}{{\mathbf{L}}}
\newcommand{\bfLp}{{\mathbf{L^{\prime}}}}
\newcommand{\dirac}{{\rm D}}

\newcommand{\pp}{{\phi\phi}}

\newlength{\tskip}
\newlength{\colwidth}

\newcommand{\beq}{\begin{equation}}
\newcommand{\eeq}{\end{equation}}
\newcommand{\beqa}{\begin{eqnarray}}
\newcommand{\eeqa}{\end{eqnarray}}
\newcommand{\bn}{\hat{\bf n}}

\newcommand{\len}{\phi}
\newcommand{\Pest}{\widehat C_{L}^{\phi\phi}}
\newcommand{\Pestp}{\widehat C_{L'}^{\phi\phi}}
\newcommand{\est}{{\mathbf{d}}_{\Theta\Theta}}
\newcommand{\estEB}{{\mathbf{d}}_{EB}}
\newcommand{\estXX}{{\mathbf{d}}_{XX}}
\newcommand{\estXXp}{{\mathbf{d}}_{XX'}}
\newcommand{\norm}{A_{\Theta\Theta}}
\newcommand{\normXX}{A_{XX}}
\newcommand{\normXXp}{A_{XX'}}
\newcommand{\filt}{F_{\Theta\Theta}}
\newcommand{\ffact}{f_{\Theta\Theta}}
\newcommand{\filtXX}{F_{XX}}
\newcommand{\filtXXp}{F_{XX'}}
\newcommand{\noise}{N^{(0)}_{\Theta\Theta,\Theta\Theta}}
\newcommand{\noiseEB}{N^{(0)}_{EB,EB}}
\newcommand{\noiseP}{N^{(1)}_{\Theta\Theta,\Theta\Theta}}
\newcommand{\noisePEB}{N^{(1)}_{EB,EB}}
\newcommand{\noiseXX}{N^{(0)}_{XX,XX}}
\newcommand{\noiseXXp}{N^{(0)}_{XX',XX'}}
\newcommand{\noiseXXpP}{N^{(1)}_{XX',XX'}}
\newcommand{\noiseXXP}{N^{(1)}_{XX,XX}}

\begin{document}
%@twocolumnfalse\endcsname

\title{Lensing Reconstruction with CMB Temperature and Polarization}
\author{Michael Kesden, Asantha Cooray, and Marc Kamionkowski}
\address{
Theoretical Astrophysics, California Institute of Technology,
Pasadena, California 91125\\}

%\date{To be submitted to Phys. Rev. D. --- October 2001}

\maketitle

%------------------------------------------------------------------------------

\begin{abstract}
 Weak gravitational lensing by intervening large-scale structure induces
 a distinct signature in the cosmic microwave background (CMB) that can be used
 to reconstruct the weak-lensing displacement map.  Estimators for individual
 Fourier modes of this map can be combined to produce an
 estimator for the lensing-potenial power spectrum.  The naive estimator for
 this quantity will be biased upwards by the uncertainty associated with
 reconstructing individual modes; we present an iterative scheme for
 removing this bias.  The variance and covariance of the lensing-potenial power
 spectrum estimator are calculated and evaluated numerically in a
 $\Lambda$CDM universe for Planck and future polarization-sensitive CMB
 experiments.
\end{abstract}

%]

%------------------------------------------------------------------------------
% User-supplied List of keywords.

%\pacs{PACS numbers: 98.80.Es,95.85.Nv,98.35.Ce,98.70.Vc
%\hfill}
%]

%------------------------------------------------------------------------------

\section{Introduction}
\label{S:intro}

The primordial cosmic microwave background (CMB) was generated when photons
first decoupled from the baryonic fluid when the universe was only 400,000
years old.  The vast majority of these photons travel unperturbed to the
present day, and features of their angular power spectrum such as acoustic
peaks and the damping tail \cite{PeeYu70} record valuable information about
cosmological parameters \cite{Jugetal95}.  Baryons and dark matter evolve
from small inhomogeneities at decoupling into increasingly complicated
large-scale structure which can subtly perturb the observed pattern of CMB
anisotropies.  Assuming that the primordial CMB is Gaussian, non-Gaussian
correlations in the observed map can be used to reconstruct the
intervening large-scale structure \cite{SelZal99}.  In addition to the
importance of learning about the large-scale structure itself, reconstruction
of the weak-lensing potential generated by structure is essential to
constraining tensor perturbations.  Weak lensing converts a fraction of the
E-mode polarization generated by scalar perturbations at the last-scattering
surface into B-mode polarization in the observed map.  Only by subtracting this
B-mode polarization can one conclusively detect the primordial
B-modes which serve as a model-independent signal of tensor perturbations
\cite{Kesetal02}.  Understanding lensing reconstruction requires a more
detailed discussion of how weak lensing affects the CMB.

Weak gravitational lensing deflects the paths of CMB photons as they travel
from the last-scattering surface to the observer.
This deflection is accomplished by a projected lensing potential which is a
weighted line-of-sight integral of the gravitational potential between the
observer and the surface of last-scattering.  At each point on the sky,
lensing remaps the temperature and polarization to that of a nearby point at
the last-scattering surface, the deflection angle being the gradient of the
aforementioned projected lensing potential.  Assuming that this deflection
angle is small, the temperature at any point can be expanded in a Taylor
series in the gradient of the lensing potential.  In Fourier space, this
expansion appears as a series of convolutions of individual temperature and
projected potential modes.  The observed temperature-squared map in Fourier
space also appears as a convolution of individual Fourier modes.  Subject to
an overall normalization dependent on the scale of the Fourier mode, these
convolutions cancel in such a manner that each Fourier mode of the
temperature-squared map acts as an estimator for the {\it same} Fourier mode
of the projected lensing potential.

Lensing reconstruction as outlined above has been considered previously
\cite{Hu01b,HuOka01}.  In these works, two sources of noise were identified,
and a filter of the temperature-squared map in Fourier space was chosen to
minimize the variance associated with lensing reconstruction subject to these
noise sources.  The first source is intrinsic signal variance; the observed
large-scale structure is one arbitrary member of an ensemble of realizations
allowed by theory.  The second source of noise, endemic to this method of
lensing reconstruction, is a consequence of the nature of the primordial CMB.
Like the large-scale structure itself, the pattern of CMB anisotropies at the
last-scattering surface is only one of many possible realizations allowed by
theory.  We do not know {\it a priori} which of these realizations nature has
provided us, and this uncertainty hinders our ability to deconvolve the effects
of lensing from true anisotropies at the last-scattering surface.  Even if the
true pattern of anisotropies at the last-scattering surface was known, the
finite amount of power in the CMB at small scales would still constrain
lensing reconstruction.  Silk damping at the last-scattering surface suppresses
CMB power at small scales, while the finite resolution of any real experiment
would limit the detection of any signal that is present at small scales.
Lensing reconstruction fails below scales at which there is sufficient power,
for the same reason that any remapping is indistinguishable given a uniform
background.

Here, we consider a third source of noise neglected in previous studies.  The
filtered temperature-squared map is an unbiased estimator for the lensing
potential in the approximation that a correlation between two given temperature
modes is induced only by the single lensing mode whose wavevector is the sum of
that of the two temperature modes.  In actuality, any combination of two or
more lensing modes whose wavevectors sum to this total induce correlations
between the two temperature modes.  There are many such combinations, but
since these correlations add incoherently we do not expect a systematic bias.
Nonetheless, for estimators of each individual lensing mode we must use our
knowledge of other lensing modes to subtract off this unwanted bias.  This is
an iterative process, and since our knowledge of the lensing map is imperfect
it induces noise in lensing reconstruction.
We calculate this additional variance for various
estimators constructed from CMB temperature and polarization maps, and show
how it compares to the dominant noise sources for an all-sky CMB experiment
with a noise-equivalent temperature of 1 $\mu$K $\sqrt{\rm sec}$.  Since the
lensing-potential power spectrum is a measure of the theoretical uncertainty
with which we can predict the value of a given lensing mode, this noise
associated with lensing reconstruction causes a systematic overestimation of
the lensing-potential power spectrum.  This systematic bias must be accounted
for in order to compare observations with theoretical predictions.

This paper is organized as follows.  In \S~\ref{S:lensing}, we define the
formalism we will use to explore the effects of weak lensing on the CMB.
The Taylor expansion of the lensed CMB map in gradients of the lensing
potential is given in both real and Fourier space, and the power spectra and
trispectra of various components of the CMB temperature map are listed for
later use.  In \S~\ref{S:estimators}, we show that the Fourier modes of the
temperature-squared map when properly filtered can serve as estimators for the
Fourier modes of the displacement map with the same wave vector.  Using
the power spectrum and trispectrum given in the preceding Section, we
calculate the variance associated with this estimator including a new
component neglected in previous studies.  This variance is evaluated
numerically using the currently favored $\Lambda$CDM cosmological model with
baryon density $\Omega_b=0.05$, matter density $\Omega_m=0.35$, cosmological
constant density $\Omega_\Lambda=0.65$, Hubble parameter $h$=0.65, and
power-spectrum amplitude $\sigma_8 = 0.9$.  We then use the displacement
estimator for individual Fourier modes to construct an unbiased estimator for
the lensing-potential power spectrum in \S~\ref{S:power}, and calculate the
variance and covariance associated with this estimator.  A few concluding
remarks about the implications of our work for future studies are given in
\S~\ref{S:disc}.  The Appendix contains useful formulae related
to additional estimators of lensing based on polarization and a
combination of temperature and polarization. 

\section{Weak Lensing of the CMB}
\label{S:lensing}

We consider weak lensing under the flat-sky approximation following
Refs.~\cite{Zal00,Hu00}.
As discussed before \cite{Hu00,SpeGol99}, weak lensing deflects the
path of CMB photons resulting in a remapping of the observed temperature
pattern on the sky,
\begin{eqnarray} \label{E:RealTay}
\tilde \cmb(\bn) & = &  \cmb[\bn + \nabla\len(\bn)] \nonumber\\
        & \approx &
\cmb(\bn) + \nabla_i \len(\bn) \nabla^i \cmb(\bn) + \frac{1}{2} \nabla_i \len(\bn) \nabla_j \len(\bn)
\nabla^{i}\nabla^{j} \cmb(\bn)
+ \ldots
\end{eqnarray}
where
$\cmb(\bn)$ is the unlensed primary component of the CMB in a direction $\bn$
at the last scattering surface.  The observed, gravitationally-lensed
temperature map $\tilde \cmb(\bn)$ in direction $\bn$ is that of the unlensed
map in direction $\bn + \nabla\len(\bn)$ where $\nabla\len(\bn)$ represents the
lensing deflection angle or displacement map.  Although a real CMB map will
include secondary contributions such as the SZ effect \cite{SunZel80}, we
assume that such effects can be distinguished by their frequency dependence
\cite{Cooetal00}.  They will not be further considered in this paper.
A noise component denoted by $\cmb^\n(\bn)$ due to finite
experimental sensitivity must be included as well.
Thus the total observed CMB anisotropy will be
$\cmb^\tot(\bn) = \tilde \cmb(\bn) + \cmb^\n(\bn)$. 

Taking  the Fourier transform of the lensed map $\tilde \cmb(\bn)$ under the
flat-sky approximation, we write
\begin{eqnarray}
\tilde \cmb(\bfl)
&=& \int d \bn\, \tilde \cmb(\bn) e^{-i \bfl \cdot \bn} \nonumber\\
&=& \cmb(\bfl) - \intlnp \cmb(\bflp) L(\bfl,\bflp)\,,
\label{E:thetal}
\end{eqnarray}
where
\begin{eqnarray}
\label{E:lfactor}
L(\bfl,\bflp) &\equiv& \len(\bfl-\bflp) \, \left[ (\bfl-\bflp) \cdot \bflp
\right]
+\frac{1}{2} \intlnpp \len(\bflpp) \\ &&\quad
\times \len(\bfl - \bflp - \bflpp) \, (\bflpp \cdot \bflp)
%\nonumber\\ &\times&
                \left[ (\bflpp + \bflp - \bfl)\cdot
                             \bflp \right] + \ldots \,.  \nonumber
\end{eqnarray}

CMB correlations in Fourier space can be described in terms of a power spectrum
and trispectrum as defined in the usual manner,
\begin{eqnarray} \label{E:specdef}
\left< \cmb^i(\bfl_1) \cmb^i(\bfl_2)\right> &\equiv&
        (2\pi)^2 \delta_\dirac(\vecl_1+\vecl_2)  C_l^i\,,\nonumber\\
\left< \cmb^i(\bfl_1) \ldots
       \cmb^i(\bfl_4)\right>_c &\equiv& (2\pi)^2 \delta_\dirac(\vecl_1+\vecl_2+\vecl_3+\vecl_4)
        T^i(\bfl_1,\bfl_2,\bfl_3,\bfl_4)\,, \nonumber \\
\end{eqnarray}
where the angle brackets denote ensemble averages over possible realizations
of the primordial CMB, large-scale structure between the observer and the
surface of last-scattering, and instrumental noise.
The subscript $c$ denotes the connected part of the four-point
function and the superscript $i$ denotes the temperature map being considered
($\cmb^\tot, \tilde \cmb,$ or $\cmb^\n$).  The lensing-potential
power spectrum can be defined analogously,
\begin{equation} \label{E:lenpowdef}
\langle \phi(\vecl) \phi(\vecl') \rangle_{\rm{LSS}}
 = (2\pi)^2 \delta_D(\vecl+\vecl') C_l^\pp \, ,
\end{equation}
where here the angle brackets denote an average over all realizations of the
large-scale structure.
We make the assumption that primordial fluctuations at the last-scattering
surface are Gaussian.  Gaussian statistics are fully described by a power
spectrum; the Gaussian four-point correlator,
$\left< \cmb(\bfl_1) \ldots \cmb(\bfl_4)\right>_c$ is zero.  The instrumental
noise $\cmb^\n$ is also assumed to be Gaussian, as is the lensing potential
$\len$.  This second assumption is justified because the dominant contributions
to the lensing potential come from intermediate redshifts
$1 \lesssim z \lesssim 3$ at which linear theory holds.

Using these definitions, we can calculate the anticipated power
spectrum and trispectrum of the observed CMB map.  Because the instrumental
noise is uncorrelated with the signal, the power spectrum of the observed map
is the sum of signal and noise power spectra,
\begin{equation} \label{E:temptot}
C_l^{\cmb\cmb\tot} = \tilde C_l^{\cmb\cmb} + C_l^{\cmb\cmb\n} \, .
\end{equation}
The power spectrum of the noise component is given by:
\begin{eqnarray} \label{E:noise}
C_l^{\cmb\cmb\n} = f_{\rm sky} w^{-1} e^{l^2 \sigma_{b}^2} \, ,
\end{eqnarray}
where $f_{\rm sky}$ is the fraction of the sky surveyed, $w^{-1}$ is the
variance per unit area on the sky, and $\sigma_b = \theta/\sqrt{8 \ln 2}$ is
the effective beamwidth of the instrument expressed in terms of its full-width
half-maximum resolution $\theta$.  A CMB experiment that spends a time
$t_{\rm pix}$ examining each of $N_{\rm pix}$ pixels with detectors of
sensitivity $s$ will have a variance per unit area
$w^{-1} = 4\pi (s/T_{\rm CMB})^2/(t_{\rm pix}N_{\rm pix})$ \cite{Kno95}.  The
power spectrum of the lensed CMB can be determined by inserting
Eq.~(\ref{E:thetal}) into Eq.~(\ref{E:specdef}) as discussed in \cite{Hu00},
\begin{eqnarray}
\tilde C_l^{\cmb\cmb} &=& \left[ 1 - \intl{1}
C^{\phi\phi}_{l_1} \left(\vecl_1 \cdot \vecl\right)^2 \right]	\, 
 	                        C_l^{\cmb\cmb}
        + \intl{1} C_{| \vecl - \vecl_1|}^{\cmb\cmb} C^{\phi\phi}_{l_1}
                [(\vecl - \vecl_1)\cdot \vecl_1]^2  \, .
\label{E:lenpow}
\end{eqnarray}
This result is given to linear order in the lensing-potential power spectrum
$C^{\phi\phi}_l$.  Lensing neither creates nor destroys power in the CMB, but
merely shifts the scales on which it occurs as seen by the fact that
\begin{equation}
\tilde \sigma^2 = \intln \tilde C_l^{\cmb\cmb} = \intln C_l^{\cmb\cmb}
= \sigma^2 \, .
\end{equation}
The observed CMB trispectrum can be calculated in a similar manner; under our
assumptions of Gaussian instrumental noise and no secondary anisotropies the
trispectrum of the lensed component $\tilde \cmb$ is the sole contribution to
the total observed trispectrum,
\begin{eqnarray}
T^\tot(\bfl_1,\bfl_2,\bfl_3,\bfl_4) &=& \tilde T^\cmb(\bfl_1,\bfl_2,\bfl_3,\bfl_4) \nonumber \\
&=& -C_{l_3}^{\cmb\cmb} C_{l_4}^{\cmb\cmb} \Big[ C^\pp_{|\vecl_1+\vecl_3|}
[(\vecla +\veclc) \cdot \veclc] [(\vecla + \veclc) \cdot \vecld]  
 + C^\pp_{|\vecl_2+\vecl_3|} [(\veclb +\veclc) \cdot \veclc] [(\veclb +\veclc)
\cdot \vecld] \Big] + \, {\rm Perm.} \, . \nonumber \\ 
\label{E:trilens}
\end{eqnarray}
The term shown above is manifestly symmetric under the interchange
$\bfl_1 \leftrightarrow \bfl_2$, while the ``+ Perm.'' represents five
additional terms identical in form but with the replacement of
$(\bfl_1, \bfl_2)$ and $(\bfl_3, \bfl_4)$ with the other five combinations of
pairs.  The total trispectrum is symmetric under the interchange of any given
pair as one would expect.  Having established a formalism within which to
analyze weak lensing, we now consider the problem of reconstructing the
lensing potential from an observed CMB temperature map.

\section{Lensing-Potential Estimators}
\label{S:estimators}

\begin{figure}[t]
%\centerline{\psfig{file=2quad.ps,width=7.0in,angle=0}}
\begin{center}
\epsfig{file=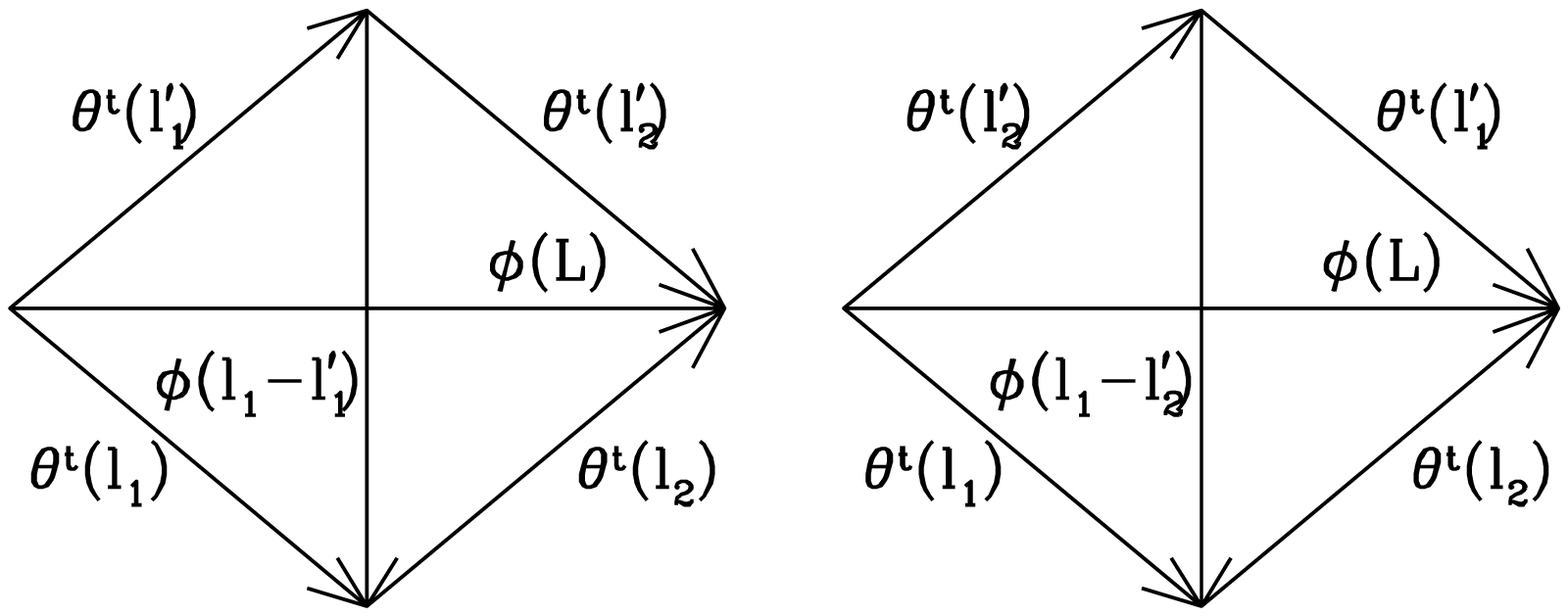,height=6.0cm,bbllx=50,bblly=350,bburx=550,bbury=570,clip=}
\caption{The two quadrilaterals consistent with the constraint
$\bfL - \bfLp = \vecla + \veclb - \vecla' - \veclb' = 0$ for the variance of
the estimator $\est(\bfL)$.  The lensing modes $\len(\bfL)$,
$\len(\vecla-\vecla')$, and $\len(\vecla-\veclb')$, depicted as diagonals in
the above quadrilaterals, induce non-Gaussian couplings between the modes of
the observed temperature map represented as sides of these quadrilaterals.
They lead to the three groups of linear terms appearing in
Eq.~(\ref{E:4pt,CMB}).
}
\label{F:2quad}
\end{center}
\end{figure}

In this Section, we examine lensing reconstruction following the approach of
Ref.~\cite{HuOka01}, largely adopting their notation as well.
The only important difference in notation is that we use $\tilde \cmb$ to
denote the lensed temperature field and $\cmb$ for the unlensed field
following Ref.~\cite{Hu00} and most recent papers.  Ref.~\cite{HuOka01} uses
the opposite convention.  For $\bfl \ne -\bflp$ and to linear order in $\len$,
\begin{equation} \label{E:EA}
\langle \cmb^\tot(\bfl) \cmb^\tot(\bflp) \rangle_{\rm{CMB}} =
f_{\cmb\cmb}(\bfl, \bflp) \phi(\bfL) \, ,
\end{equation}
where
\begin{equation} \label{E:bias}
f_{\cmb\cmb}(\bfl, \bflp) \equiv C_{l}^{\cmb\cmb} (\bfL \cdot \bfl) +
C_{l^{\prime}}^{\cmb\cmb} (\bfL \cdot \bflp) \, ,
\end{equation}
and $\bfL = \bfl + \bflp$.
Note that $\langle \quad \rangle_{\rm{CMB}}$ differs from the unmarked
$\langle \quad \rangle$ that first appeared in Eq.~(\ref{E:specdef}) in that
it denotes an ensemble average only over different
Gaussian realizations of the primordial CMB and instrument noise; a fixed
realization of the large-scale structure is assumed.  For the purposes of
estimating the large-scale structure actually realized in our observable
universe, this is the appropriate average to take to ensure that our
estimators are truly unbiased for a typical realization of the primordial CMB.
When calculating the noise associated with lensing-potential estimators and
again for power spectrum estimation in \S~\ref{S:power}, we will return to
the full unmarked ensemble average. Eq.~(\ref{E:EA}), 
an immediate consquence of Eq.~(\ref{E:thetal}), suggests that a
temperature-squared map appropriately filtered in Fourier space can serve
as an estimator for the deflection field $\bfd(\bfL) \equiv i\bfL \len(\bfL)$.
Hu and Okamoto define five different estimators for the deflection field
constructed from various combinations of the temperature and polarization; we
discuss the temperature-squared estimator in this Section and relegate the
analogous formulae for polarization estimators to the Appendix.  The
minimum-variance temperature-squared estimator derived in Ref.~\cite{HuOka01}
is
\begin{equation} \label{E:est}
\est(\bfL) \equiv \frac{i\bfL \norm(L)}{L^2} \intl{1} \cmb^\tot(\vecla)
\cmb^\tot(\veclb) \filt(\vecla, \veclb) \, ,
\end{equation}
where
\begin{equation} \label{E:filt}
\filt(\vecla, \veclb) \equiv \frac{f_{\cmb\cmb}(\vecla, \veclb)}
{2 C_{l_1}^{\cmb\cmb\tot} C_{l_2}^{\cmb\cmb\tot}} \, ,
\end{equation}
\begin{equation} \label{E:norm}
\norm(L) \equiv L^2 \left[ \intl{1} f_{\cmb\cmb}(\vecla, \veclb)
\filt(\vecla, \veclb) \right]^{-1} \, ,
\end{equation}
and $\veclb = \bfL - \vecla$.  Substitution of Eqs.~(\ref{E:EA}),
(\ref{E:bias}), (\ref{E:filt}), and (\ref{E:norm}) into Eq.~(\ref{E:est}) shows
the desired result,
\begin{equation}
\langle \est(\bfL) \rangle_{\rm{CMB}} = \bfd(\bfL) \, ,
\end{equation}
namely that $\est(\bfL)$ is indeed an unbiased estimator for the deflection
field in Fourier space.  We now proceed to calculate the variance of this
estimator.  At first, we assume a complete knowledge of all lensing modes not
examined by this estimator.  In that case, we find
\begin{eqnarray} \label{E:estvar,CMB}
&& \langle \est^{\ast}(\bfL) \cdot \est(\bfLp) \rangle_{\rm{CMB}} -
\langle \est^{\ast}(\bfL) \rangle_{\rm{CMB}} \cdot
\langle \est(\bfLp) \rangle_{\rm{CMB}}
=
(\bfL \cdot \bfLp) \frac{\norm(L)}{L^2}
\frac{\norm(L^{\prime})}{L^{\prime2}} \nonumber \\
&& \quad \quad \times \intl{1} \intl{1'} \langle
\cmb^\tot(-\vecla) \cmb^\tot(-\veclb) \cmb^\tot(\vecla') \cmb^\tot(\veclb')
\rangle_{\rm{CMB}}
\filt(\vecla, \veclb) \filt(\vecla', \veclb')
- \bfd^{\ast}(\bfL) \cdot \bfd(\bfLp)
\, , \nonumber \\
\end{eqnarray}
where $\veclb' = \bfLp - \vecla'$.  Evaluating the four-point function in the
integrand of Eq.~(\ref{E:estvar,CMB}) to second order in the lensing field,
we obtain
\begin{eqnarray} \label{E:4pt,CMB}
&& \langle
\cmb^\tot(-\vecla) \cmb^\tot(-\veclb) \cmb^\tot(\vecla') \cmb^\tot(\veclb')
\rangle_{\rm{CMB}} = \nonumber \\
&& \quad
\biggl[ \left( C_{l_1}^{\cmb\cmb} +
C_{l_1}^{\cmb\cmb\n} \right) (2 \pi)^2 \delta_\dirac(\bfL) + \len(-\bfL)
\ffact(\vecla, \veclb) - \intlnp C_{l^{\prime}}^{\cmb\cmb}
\len(-\vecla - \bflp) \len(-\veclb + \bflp) \bigl[ \bflp \cdot (\vecla + \bflp)
\bigr] \nonumber \\
&& \quad \times \bigl[ \bflp \cdot (\veclb - \bflp) \bigr] -
\frac{1}{2} \intlnp \len(\bflp) \len(-\bfL - \bflp) 
\Bigl\{ C_{l_1}^{\cmb\cmb} (\vecla \cdot \bflp) \bigl[ \vecla \cdot
(\bfL + \bflp) \bigr] + C_{l_2}^{\cmb\cmb} (\veclb \cdot \bflp)
\bigl[ \veclb \cdot (\bfL + \bflp) \bigr]
\Bigr\} \biggr] \nonumber \\
&& \quad \times
\biggl[ \bigl( C_{l_{1}^{\prime}}^{\cmb\cmb} +
C_{l_{1}^{\prime}}^{\cmb\cmb\n} \bigr) (2 \pi)^2 \delta_\dirac(\bfLp) +
\len(\bfLp) \ffact(\vecla', \veclb') - \intlnp C_{l^{\prime}}^{\cmb\cmb}
\len(\vecla' - \bflp) \len(\veclb' + \bflp) \bigl[ \bflp \cdot
(\vecla' - \bflp) \bigr] \nonumber \\
&& \quad \times \bigl[ \bflp \cdot (\veclb' + \bflp) \bigr] +
\frac{1}{2} \intlnp \len(\bflp) \len(\bfLp - \bflp)
\Bigl\{ C_{l_{1}^{\prime}}^{\cmb\cmb} (\vecla' \cdot \bflp) \bigl[ \vecla'
\cdot (\bflp - \bfLp) \bigr] + C_{l_{2}^{\prime}}^{\cmb\cmb} (\veclb' \cdot
\bflp) \bigl[ \veclb' \cdot (\bflp - \bfLp) \bigr] \Bigr\}
\biggr] {\rm + Perm.} \nonumber \\
\end{eqnarray}
The terms given explicitly in Eq.~(\ref{E:4pt,CMB}) correspond to the
correlations between $\cmb^\tot(-\vecla)$ and $\cmb^\tot(-\veclb)$ and
those between $\cmb^\tot(\vecla')$ and $\cmb^\tot(\veclb')$.
The ``+ Perm.'' stands for two additional terms, identical in form,  arising
from the pairings $\langle \cmb^\tot(-\vecla) \cmb^\tot(\vecla')
\rangle_{\rm{CMB}} \langle \cmb^\tot(-\veclb) \cmb^\tot(\veclb')
\rangle_{\rm{CMB}}$ and $\langle \cmb^\tot(-\vecla) \cmb^\tot(\veclb')
\rangle_{\rm{CMB}} \langle \cmb^\tot(-\veclb) \cmb^\tot(\vecla')
\rangle_{\rm{CMB}}$.
This expression indicates how uncertainty in the CMB at the last-scattering
surface propagates into uncertainty in lensing reconstruction for a particular
realization $\len(\bfL)$ of the large-scale structure.  To linear order in
Eq.~(\ref{E:4pt,CMB}), correlations between the modes
$\cmb^\tot(-\vecla), \cmb^\tot(-\veclb), \cmb^\tot(\vecla'),$ and
$\cmb^\tot(\vecla')$ are induced by those lensing modes whose wavevectors are
the sums of any pair of wavevectors of these four modes.  These lensing modes
are precisely those forming the diagonals of the quadrilaterals depicted in
Fig.~\ref{F:2quad}.
In practice, we do not
know the large-scale structure between us and the last-scattering surface, so
we assume a variance given by Eq.~(\ref{E:lenpowdef}) with a model-dependent
power spectrum $C_L^\pp$.  We must average Eq.~(\ref{E:estvar,CMB})
over different realizations of the large-scale structure (denoted by
$\langle \quad \rangle_{\rm{LSS}}$) to obtain the total expected variance of
our estimator,
\begin{eqnarray} \label{E:estvar}
&& \Big\langle \langle \est^{\ast}(\bfL) \cdot \est(\bfLp) \rangle_{\rm{CMB}} -
\langle \est^{\ast}(\bfL) \rangle_{\rm{CMB}} \cdot
\langle \est(\bfLp) \rangle_{\rm{CMB}} \Big\rangle_{\rm{LSS}}
=
(\bfL \cdot \bfLp) \frac{\norm(L)}{L^2}
\frac{\norm(L^{\prime})}{L^{\prime2}} \nonumber \\
&& \quad \quad \times \intl{1} \intl{1'} \langle
\cmb^\tot(-\vecla) \cmb^\tot(-\veclb) \cmb^\tot(\vecla') \cmb^\tot(\veclb')
\rangle
\filt(\vecla, \veclb) \filt(\vecla', \veclb')
- (2 \pi)^2 \delta_\dirac(\bfL - \bfLp) C_{L}^{dd}
\, , \nonumber \\
\end{eqnarray}
where
\begin{equation}
\langle \bfd^{\ast}(\bfL) \cdot \bfd(\bfLp) \rangle = (2 \pi)^2
\delta_\dirac(\bfL - \bfLp) C_{L}^{dd} =
(2 \pi)^2 \delta_\dirac(\bfL - \bfLp) L^2 C_{L}^{\len\len} \, .
\end{equation}
The assumption that the lensing potential is Gaussian imposes the constraint
$\bfL - \bfLp = \vecla + \veclb - \vecla' - \veclb' = 0$ which closes the
quadrilaterals of Fig.~\ref{F:2quad}.
The average of the four-point correlation function in Eq.~(\ref{E:estvar}) can
be calculated by further averaging Eq.~(\ref{E:4pt,CMB}) over the large-scale
structure.  Terms linear in the lensing field vanish when averaged over
different realizations of the large-scale structure.  Quadratic terms in the
lensing field arise as products either of two linear terms or of a zeroth and
second-order term.  Averages over the product of two linear terms produce the
connected part of the four-point correlation function, the trispectrum defined
in Eq.~(\ref{E:specdef}).  Averages over the product of a zeroth and
second-order term have no connected portion, but instead furnish an implicit
dependence on $C_L^\pp$ in the total observed power spectrum of
Eq.~(\ref{E:temptot}).  The final result of averaging over the large-scale
structure can be expressed
in terms of the observed power spectrum and trispectrum,
\begin{eqnarray} \label{E:4pt}
\langle
\cmb^\tot(-\vecla) \cmb^\tot(-\veclb) \cmb^\tot(\vecla') \cmb^\tot(\veclb')
\rangle &=& (2 \pi)^4 \Big[ C_{l_1}^{\cmb\cmb\tot}
C_{l_{1}^{\prime}}^{\cmb\cmb\tot}
\delta_\dirac(\bfL) \delta_\dirac(\bfLp) + C_{l_1}^{\cmb\cmb\tot}
C_{l_2}^{\cmb\cmb\tot} \Big\{
\delta_\dirac(\vecla' - \vecla) \delta_\dirac(\veclb' - \veclb)
\nonumber \\ && \quad
+ \delta_\dirac(\veclb' - \vecla) \delta_\dirac(\vecla' - \veclb) \Big\}
+ (2 \pi)^{-2}
T^\tot(-\vecla,-\veclb,\vecla',\veclb') \delta_\dirac(\bfL - \bfLp) \Big]
\, , \nonumber \\
\end{eqnarray}
where the trispectrum can written in terms of $\ffact(\vecla,\veclb)$ as
\begin{equation}
T^\tot(\bfl_1,\bfl_2,\bfl_3,\bfl_4) = C_{|\vecla+\veclb|}^\pp
\ffact(\vecla,\veclb) \ffact(\veclc,\vecld) + C_{|\vecla+\veclc|}^\pp
\ffact(\vecla,\veclc) \ffact(\veclb,\vecld) + C_{|\vecla+\vecld|}^\pp
\ffact(\vecla,\vecld) \ffact(\veclb,\veclc) \, .
\end{equation}
This form of the trispectrum is consistent with that of
Eq.~(\ref{E:trilens}) given directly in terms of power spectra.
Since $\bfL \ne 0$, $\delta_\dirac(\bfL) = 0$ and the first term of
Eq.~(\ref{E:4pt}) vanishes.  The remaining two terms
containing pairs of delta functions, inserted into Eq.~(\ref{E:estvar}), yield
the dominant contribution to the variance,
\begin{equation}
\Big\langle \langle \est^{\ast}(\bfL) \cdot \est(\bfLp) \rangle_{\rm{CMB}} -
\langle \est^{\ast}(\bfL) \rangle_{\rm{CMB}} \cdot
\langle \est(\bfLp) \rangle_{\rm{CMB}} \Big\rangle_{\rm{LSS}}
= (2 \pi)^2
\delta_\dirac(\bfL - \bfLp) [ \noise(L) + \ldots \, ] \, ,
\end{equation}
where $\noise(L) = \norm(L)$.  Notice that $\noise(L)$ is
zeroth order in the lensing potential $\len$; it depends on the lensing
potential power spectrum $C_{l}^{\len\len}$ only implicitly though the total
observed power spectrum $C_{l}^{\cmb\cmb\tot}$.  The ellipsis represents terms
of higher order in $C_{l}^{\len\len}$ that we now proceed to calculate.  These
terms arise from the trispectrum term of Eq.~(\ref{E:4pt}) after
$\langle  \langle \est^{\ast}(\bfL) \rangle_{\rm{CMB}} \cdot
\langle \est(\bfLp) \rangle_{\rm{CMB}} \rangle_{\rm{LSS}} =
(2 \pi)^2 \delta_\dirac(\bfL - \bfLp) C_{L}^{dd}$ is removed.  Substituting
these results into Eq.~(\ref{E:estvar}),
we find that to first order in $C_{l}^{\len\len}$,
\begin{equation} \label{E:prodave}
\Big\langle \langle \est^{\ast}(\bfL) \cdot \est(\bfLp) \rangle_{\rm{CMB}} -
\langle \est^{\ast}(\bfL) \rangle_{\rm{CMB}} \cdot
\langle \est(\bfLp) \rangle_{\rm{CMB}} \Big\rangle_{\rm{LSS}}
= (2 \pi)^2
\delta_\dirac(\bfL - \bfLp) [ \noise(L) + \noiseP(L) ] \, ,
\end{equation}
where $\noiseP(L)$ is given by,
\begin{eqnarray} \label{E:1st}
&& \noiseP(L) =  \frac{\norm^2(L)}{L^2} \intl{1} \intl{1'}
\filt(\vecla, \veclb) \filt(\vecla', \veclb') \nonumber \\
&& \quad \times \Big\{
C_{|\vecla - \vecla'|}^{\phi\phi} f_{\cmb\cmb}(-\vecla, \vecla')
f_{\cmb\cmb}(-\veclb, \veclb') +
C_{|\vecla-\veclb'|}^{\phi\phi} f_{\cmb\cmb}(-\vecla, \veclb')
f_{\cmb\cmb}(-\veclb, \vecla') \Big\} \, . \nonumber \\
\end{eqnarray}
The first-order contribution to
the noise $\noiseP(L)$ involves integrals over the lensing-potential power
spectrum, and thus probes lensing modes with wave vectors different from that
of the estimator $\est(\bfL)$.  It can be interpreted physically as
interference from these other modes in the determination of the mode
$\bfd(\bfL)$ being estimated.
The filter $\filt(\vecla, \veclb)$ was chosen to optimize the
signal-to-noise ratio in the absence of the first-order contribution
$\noiseP(L)$; it is no longer an optimal filter once this additional
noise is taken into account.  As long as $\noiseP(L) \ll \noise(L)$, the noise
reduction that can be attained by re-optimizing our filter will not be
significant.  Formulas analagous to those presented here relevant to the
construction of estimators using polarization data are given in the Appendix.

\begin{figure}[t]
%\centerline{\psfig{file=fig2.ps,width=7.0in,angle=0}}
\begin{center}
\epsfig{file=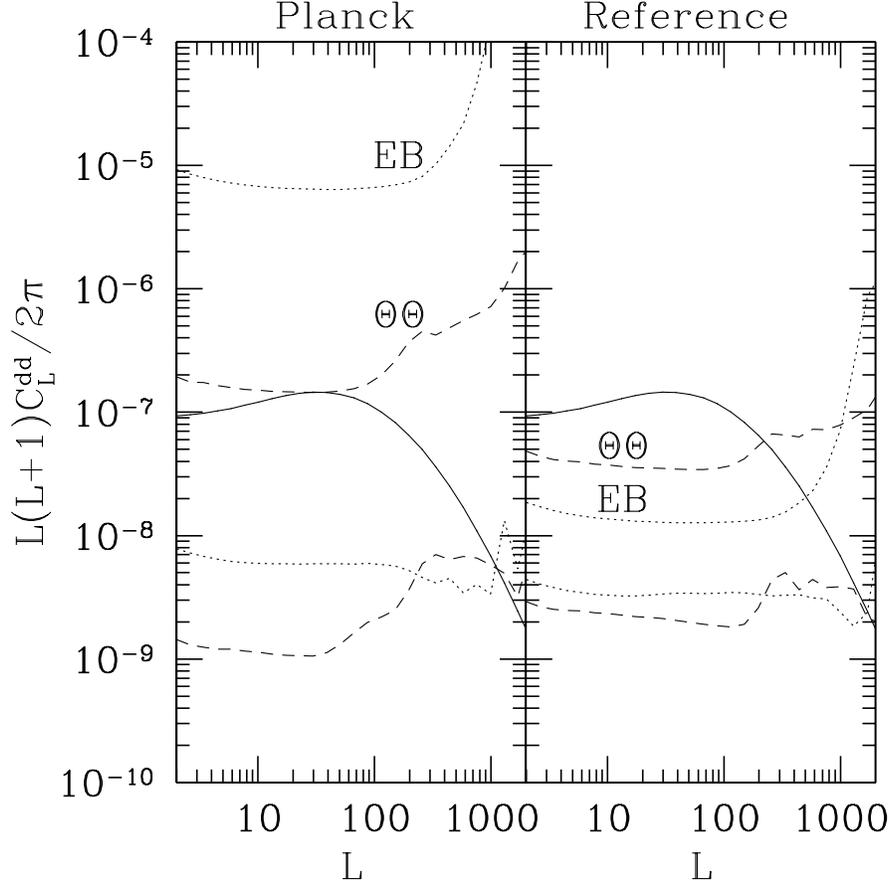,height=12cm,bbllx=20,bblly=150,bburx=570,bbury=715,clip=}
\caption{Variances with which individual modes $\bfd(\bfL)$ of the deflection
field can be reconstructed by the Planck and reference experiments
described in the text.  The solid curves are the power spectra $C_{L}^{dd}$
anticipated for our $\Lambda$CDM cosmological model.  The
upper and lower dashed curves are the zeroth and first-order noise power
spectra $\noise(L)$ and $\noiseP(L)$ respectively for the temperature-based
estimator $\est(\bfL)$, while the dotted curves are the corresponding noise
variances for $\estEB(\bfL)$.  A mode $\bfd(\bfL)$ cannot be reconstructed
with signal-to-noise greater than unity when $C_{L}^{dd} \leq \noise(L) +
\noiseP(L)$.}
\label{F:Plrec}
\end{center}
\end{figure}

The significance of $\noiseP(L)$ for two different experiments
is shown in Fig.~\ref{F:Plrec} using the currently favored $\Lambda$CDM
cosmological model with baryon density $\Omega_b=0.05$, matter density
$\Omega_m=0.35$, cosmological constant density $\Omega_\Lambda=0.65$, Hubble
parameter $h$=0.65, and power-spectrum amplitude $\sigma_8 = 0.9$.  The Planck
experiment is equivalent to a one-year, full-sky survey with temperature and
polarization sensitivities of 12.42 and 26.02 $\mu \rm{K} \sqrt{\rm{sec}}$
respectively and resolution $\theta = 7.0$
arcminutes as described in \S~\ref{S:lensing}.  The reference experiment has
the same resolution but superior sensitivities of 0.46 and 0.65 $\mu \rm{K}
\sqrt{\rm{sec}}$ for temperature and polarization.  These estimates of
experimental parameters are identical to those given for the Planck and
reference experiments of Ref.~\cite{HuOka01}.  The $\cmb E$ and
$EE$ estimators have noise power spectra intermediate to those of the
$\cmb\cmb$ and $EB$ estimators, while the $\cmb B$ estimator has substantially
higher noise because the primordial CMB lacks true B-modes in the absence of
inflationary gravitational waves.  We see that for Planck, with its
comparatively inferior polarization sensitivity, the $\cmb\cmb$ estimator will
be best although it will be unable to detect individual Fourier modes of the
deflection field at the $1\sigma$ level.  The reference experiment, and further
experiments with similar sensitivity and even higher resolution, should be able
to push $1\sigma$ detection of individual $\bfd(\bfL)$ modes to $L \simeq 1000$
by primarily relying on the $EB$ estimator.  In these cases the secondary noise
$\noisePEB(L)$ is only smaller than the dominant noise $\noiseEB(L)$ by a
factor of a few, whereas for higher sensitivity (noisier) experiments like
Planck it is smaller by at least an
order-of-magnitude.  This illustrates an interesting point, apparent from
Fig.~\ref{F:Plrec}, that the zeroth-order noise $\noise(L)$ declines
dramatically with decreasing sensitivity until it becomes dominated by cosmic
variance while the $\noiseP(L)$ is largely unaffected by instrument
sensitivity.  The reasons for this trend are that instrument noise appears in
$\noise(L) = \norm(L)$ through its contribution to the denominator of
$\filt(\vecla,\veclb)$ as shown by Eqs.~(\ref{E:filt}) and (\ref{E:norm}).
Decreasing instrument noise raises the value of $\filt(\vecla,\veclb)$ thereby
lowering $\noise(L)$.  By contrast instrument noise is reflected in
$\noiseP(L)$
through its effects on both $\norm(L)$ and $\filt(\vecla,\veclb)$ as shown in
Eq.~(\ref{E:1st}).  Smaller instrument noise raises $\filt(\vecla,\veclb)$
as before, driving $\noiseP(L)$ up in this case, but this is compensated by a
decrease in $\norm(L)$ which appears as a prefactor outside the integrals.
These two effects largely cancel each other out, rendering $\noiseP(L)$
remarkably insensitive to instrument noise.

\section{Power Spectrum Estimation}
\label{S:power}

Although complete reconstruction of the deflection field $\bfd(\bfL)$ can be
an enormously powerful tool, such as for B-mode subtraction \cite{Kesetal02},
for some purposes estimates of the lensing-potential power spectrum
$C_{L}^{\len\len}$ are sufficient.  This power spectrum is a model-dependent
prediction of theories of large-scale structure formation, and therefore
estimates of the power spectrum from real data could be used to test these
theories as well as the consistency of other determinations of cosmological
parameters.  Furthermore, since estimates of all the modes $\bfd(\bfL)$ with
$|\bfL| = L$ can be combined to estimate $C_{L}^{\len\len}$, $1\sigma$
detection of the power spectrum can be pushed to much higher $L$ than can that
of individual modes.  The deflection-field estimator $\est(\bfL)$ derived in
the preceding Section can be used to construct an estimator for
$C_{L}^{\len\len}$.  Our first guess for an appropriate lensing-potential power
spectrum estimator is
\begin{equation} \label{E:naive}
D_L \equiv \frac{(2 \pi)^2}{A L^2} \frac{1}{2 \pi L \Delta L} \int_{a_L} \VE
\, \est(\bfl) \cdot \est(-\bfl) \, ,
\end{equation}
where $A$ is the area of the sky surveyed and $a_L$ is an annulus of radius
$L$ and width $\Delta L$.  We ensemble average our estimator over different
realizations of the CMB and large-scale structure using Eq.~(\ref{E:prodave})
by bringing $\langle \langle \est^{\ast}(\bfL) \rangle_{\rm{CMB}} \cdot
\langle \est(\bfLp) \rangle_{\rm{CMB}} \rangle_{\rm{LSS}}$ from the left to
the right-hand side.  This yields
\begin{equation} \label{E:naive2}
\langle D_L \rangle = \frac{(2 \pi)^2}{A L^2} \frac{1}{2 \pi L \Delta L}
\delta_\dirac(0) \int_{a_L} d^2 \bfl \left[ l^2 C_{l}^{\len\len} +
\noise(l) + \noiseP(l) \right] \, .
\end{equation}
The definition of the Dirac delta function,
\begin{equation}
(2 \pi)^2 \delta_\dirac(\bfl) \equiv \int_{A} d \bn \, e^{i \bfl \cdot \bn}
\, ,
\end{equation}
implies that $\delta_\dirac(0) = A/(2 \pi)^2$.  Furthermore, in the limit that
$\Delta L$ is small compared to the scales on which $l^2 C_{l}^{\len\len} +
\noise(l) + \noiseP(l)$ is varying, we can evaluate the integrand of
Eq.~(\ref{E:naive2}) at its central value $l = L$ and extract it from the
integral.  The integral over the annulus $a_L$ cancels the factor $2 \pi L
\Delta L$ in the denominator, reducing Eq.~(\ref{E:naive2}) to
\begin{equation} \label{E:naive3}
\langle D_L \rangle = C_{L}^{\len\len} + L^{-2} \left[ \noise(L) + \noiseP(L)
\right] \, .
\end{equation}
$D_L$ is indeed an estimator for the lensing-potential power spectrum
$C_{L}^{\len\len}$, albeit a biaed one.  Note that the {\it bias} in the
power-spectrum estimator $D_L$ is precisely the same as the {\it variance}
shown in Eq.~(\ref{E:prodave}) with which we were able to determine each
individual lensing mode.  This is no coincidence; it reflects the fact that
there are no grounds {\it a priori} on which to differentiate the variance
with which we can reconstruct individual modes $\bfd(\bfL)$ from the intrinsic
variance $C_{L}^{dd}$ of the underlying distribution from which they are
drawn.  To obtain an unbiased estimator to compare with theoretical
predictions, we subtract off this unwanted reconstruction variance,
\begin{equation} \label{E:PestSub}
\Pest \equiv D_L - L^{-2} \left[ \noise(L) + \noiseP(L) \right] \, .
\end{equation}
Since $\noiseP(L)$ as defined in Eq.~(\ref{E:1st}) itself depends on
$C_{L}^{\len\len}$, this subtraction and evaluation must be performed
iteratively until a self-consistent solution is obtained.
The variance of our estimator $\Pest$ can be calculated in the usual manner,
\begin{equation} \label{E:sigPV}
\sigma_{\Pest}^2 \equiv \langle (\Pest)^2 \rangle - \langle \Pest \rangle^2
= \langle (D_L)^2 \rangle - \langle D_L \rangle^2 \, .
\end{equation}
Evaluating this expression requires us to calculate
\begin{equation} \label{E:DL2}
\langle (D_L)^2 \rangle = \frac{(2 \pi)^4}{A^2 L^4}
\frac{1}{(2 \pi L \Delta L)^2} \int_{a_L} \frac{d^2 \vecla'}{(2 \pi)^2} \,
\int_{a_L} \frac{d^2 \veclb'}{(2 \pi)^2} \, \langle \left[
\est(\vecla') \cdot \est(-\vecla') \right] \left[
\est(\veclb') \cdot \est(-\veclb') \right] \rangle \, .
\end{equation}
Since $\est(\bfL)$ is a quadratic estimator in the temperature map,
Eq.~(\ref{E:DL2}) includes the following integral over the eight-point
correlation function in Fourier space,
\begin{eqnarray} \label{E:8pt}
&& \langle \left[
\est(\vecla') \cdot \est(-\vecla') \right] \left[
\est(\veclb') \cdot \est(-\veclb') \right] \rangle =
\frac{\norm^2(l_{1}^{\prime}) \norm^2(l_{2}^{\prime})}{(l_{1}^{\prime})^2
(l_{2}^{\prime})^2} \intk{1} \intk{3} \intk{5} \intk{7} \nonumber \\
&& \quad \times \Big\{ \filt(\vecka, \veckb) \filt(\veckc, \veckd)
\filt(\vecke, \veckf) \filt(\veckg, \veckh) \langle \cmb^\tot(\vecka)
\cmb^\tot(\veckb) \cmb^\tot(\veckc) \cmb^\tot(\veckd) \cmb^\tot(\vecke)
\cmb^\tot(\veckf) \cmb^\tot(\veckg) \cmb^\tot(\veckh) \rangle \Big\}
\, , \nonumber \\
\end{eqnarray}
where $\veckb = \vecla' - \vecka$, $\veckd = -\vecla' - \veckc$,
$\veckf = \veclb' - \vecke$, and $\veckh = -\veclb' - \veckg$.  A fully
general eight-point correlation function consists of a connected part, as
well as terms proportional to the product of lower-order correlation
functions.  Under the assumption that both the primordial CMB and the lensing
potential are governed by Gaussian statistics, all correlation functions higher
than the four-point have vanishing connected parts \cite{KesCum}.
The temperature eight-point correlation function will therefore be composed
of three groups of terms; membership in a group being determined by whether the
term contains zero, one, or two factors of the trispectrum.
Since the trispectrum given in Eq.~(\ref{E:trilens}) is first order in the
lensing-potential power spectrum $C_{L}^{\len\len}$, terms of these three
groups are zeroth, first, and second order respectively in $C_{L}^{\len\len}$.
Combinatorics determines the number of terms in each group.  There are:
$\frac{1}{4!} \dbinom{8}{2} \dbinom{6}{2} \dbinom{4}{2} \dbinom{2}{2} = 105$
different ways of dividing $(\vecka, \ldots, \veckh)$ into four pairs, and
hence there will be 105 terms in the group containing no trispectra.
Similar calculations reveal that there are $\frac{1}{2!} \dbinom{8}{4}
\dbinom{4}{2} \dbinom{2}{2} = 210$ terms in the second group and $
\frac{1}{2!} \dbinom{8}{4} \dbinom{4}{4} = 35$ terms in the third group.  Many
terms in all three groups will vanish for the same reason that the first term
vanished in the four-point correlation function of Eq.~(\ref{E:4pt}); these
terms are proportional to a Dirac delta function evaluated at nonzero argument.
Consider now the first group of terms, those that are zeroth order in
$C_{L}^{\len\len}$.  The 60 nonvanishing terms in this group each contain
four Dirac delta functions; they can be further segregated into the 12 terms
that allow two of the integrals over $\vecki$ appearing in Eq.~(\ref{E:8pt}) to
be immediately evaluated via Dirac delta functions, and the 48 terms that
allow evaluation of three $\vecki$ integrals.  The first 12 terms, inserted
into Eq.~(\ref{E:8pt}) and appropriately evaluated using the normalization of
Eq.~(\ref{E:norm}), yield
\begin{equation}
\langle (D_L)^2 \rangle = \left( \frac{\noise(L)}{L^2} \right)^2
\left[ 1 + 2 \frac{(2 \pi)^2}{A} \frac{1}{2 \pi L \Delta L} \right]
+ \ldots \, ,
\end{equation}
while the remaining 48 terms give the final result to zeroth order in
$C_{L}^{\len\len}$,
\begin{eqnarray} \label{E:zeroPV}
&& \langle (D_L)^2 \rangle = \left( \frac{\noise(L)}{L^2} \right)^2
\left[ 1 + 2 \frac{(2 \pi)^2}{A} \frac{1}{2 \pi L \Delta L} \right] 
\nonumber \\ && \quad \quad \quad \quad \quad \quad +
\frac{(2 \pi)^2}{A L^4}
\frac{2}{(2 \pi L \Delta L)^2} \int_{a_L} \frac{d^2 \vecla'}{(2 \pi)^2} \,
\int_{a_L} \frac{d^2 \veclb'}{(2 \pi)^2} \, \frac{\norm^2(l_{1}^{\prime})
\norm^2(l_{2}^{\prime})}{(l_{1}^{\prime})^2 (l_{2}^{\prime})^2} \intk{1}
f_{\cmb\cmb}(\vecka, \veckb) P(\vecka, \veckb, \vecla', \veclb') \, ,
\nonumber \\
\end{eqnarray}
where
\begin{eqnarray}
P(\vecka, \veckb, \vecla', \veclb') &=& \ffact(-\veclb' - \vecka, \veclb' -
\veckb) \filt(-\vecka, \veclb' + \vecka) \filt(-\veckb, -\veclb' + \veckb)
\nonumber \\
&& + \ffact(\veclb' - \vecka, -\veclb' -
\veckb) \filt(-\vecka, -\veclb' + \vecka) \filt(-\veckb, \veclb' + \veckb)
\nonumber \\
&& + \ffact(-\veclb' - \vecka, \vecla' -
\veckb) \filt(-\vecka, \veclb' + \vecka) \filt(-\veckb, -\vecla' + \veckb)
\nonumber \\
&& + \ffact(\vecla' - \vecka, -\veclb' -
\veckb) \filt(-\vecka, -\vecla' + \vecka) \filt(-\veckb, \veclb' + \veckb)
\nonumber \\
&& + \ffact(\vecla' - \vecka, \veclb' -
\veckb) \filt(-\vecka, -\vecla' + \vecka) \filt(-\veckb, -\veclb' + \veckb)
\nonumber \\
&& + \ffact(\veclb' - \vecka, \vecla' -
\veckb) \filt(-\vecka, -\veclb' + \vecka) \filt(-\veckb, -\vecla' + \veckb)
\, . \nonumber \\
\end{eqnarray}
When $\langle D_L \rangle^2$ is subtracted from $\langle (D_L)^2 \rangle$ in
Eq.~(\ref{E:sigPV}), the first term in the square brackets of
Eq.~(\ref{E:zeroPV}) will be eliminated.  Minimizing the variance associated
with this estimator
then consists of making an optimal choice of $\Delta L(L)$.  The first noise
term is proportional to $\frac{(2 \pi)^2/A}{2 \pi L \Delta L}$.  For a survey
of area $A$, $(2 \pi)^2/A$ is the specific area of an individual mode in
$\bfL$ space and $2 \pi L \Delta L$ is the area in $\bfL$ space over which
the power-spectrum estimator takes an average.  This ratio is therefore the
inverse of the number of individual $\est(\bfL)$ modes whose inverse variances
are added to determine the inverse variance of $\Pest$.  It is obviously
minimized by choosing $(2 \pi)^2/A \ll 2 \pi L \Delta L(L)$.  The second term,
that involving $P(\vecka, \veckb, \vecla', \veclb')$, differs from the first
noise term in that a Dirac delta function has been used to evaluate an
additional $\vecki$ integral rather than an annulus integral.  Since the
integrands are of the same order, we expect the second noise term to be
suppressed relative to the first by a factor $2 \pi L \Delta L(L)/ \pi
l_{\rm{max}}^2$ where $l_{\rm{max}} \simeq \pi/\theta$ is set by the resolution
$\theta$ of the survey.  Under the conservative assumption
$L \Delta L(L) \ll 1/\theta^2$, namely that we are probing scales well above
our resolution, this term is assured to be small.  We neglect such terms for
the remainder of this paper.  If we insert the portions of the eight-point
correlation function that are first and second order in $C_{L}^{\len\len}$
into Eq.~(\ref{E:8pt}) and evaluate using Eq.~(\ref{E:1st}), we find
\begin{equation}
\langle (D_L)^2 \rangle = L^{-4} \left( C_{L}^{dd} + \noise(L) + \noiseP(L)
\right)^2 \left[ 1 + 2 \frac{(2 \pi)^2}{A} \frac{1}{2 \pi L \Delta L} \right]
\, ,
\end{equation}
and
\begin{equation} \label{E:FinVar}
\sigma_{\Pest}^2 = 2 \frac{(2 \pi)^2}{A} \frac{1}{2 \pi L \Delta L} L^{-4}
\left( C_{L}^{dd} + \noise(L) + \noiseP(L) \right)^2 \, .
\end{equation}
This result agrees with that given in Ref.~\cite{HuOka01} after subtracting
our newly derived term $\noiseP(L)$.

The term $\noiseP(L)$ and corresponding terms for polarization-based
estimators have two principal effects on power spectrum estimation.  As shown
in Eq.~(\ref{E:FinVar}), they provide a fractional contribution to the variance
of roughly $2\noiseP(L)/C_{L}^{dd}$ when $C_{L}^{dd}$ dominates the variance
as in the reference experiment in the right-hand panel of Fig.~\ref{F:Plrec}.
For the $\Lambda$CDM cosmological model considered here this represents an
increase of $5-15 \%$ in the variance of the $EB$ estimator for
$L \lesssim 1000$.  More importantly, $\noiseP(L)$ acts as a bias for the naive
estimator $D_L$ as shown by Eq.~(\ref{E:naive3}).  If this bias is not
calculated and subtracted iteratively to form the unbiased estimator $\Pest$ as
in Eq.~(\ref{E:PestSub}), $C_{L}^{dd}$ will be {\it systematically}
overestimated by $5-10 \%$ at low $L$ and by increasingly larger amounts at
$L \gtrsim 100$ as the signal $L(L+1)C_{L}^{dd}/2\pi$ begins to plummet while
$\noiseP(L)$ remains comparatively flat.

Having evaluated the variance of our estimator $\Pest$, we consider whether
this estimator has a substantial covariance
\begin{equation} \label{E:Pcov}
\sigma_{\Pest \Pestp} \equiv \langle (\Pest - C_{L}^{\len\len})
(\Pestp - C_{L'}^{\len\len}) \rangle
= \langle D_L D_{L'} \rangle - \langle D_L \rangle \langle D_{L'} \rangle \, .
\end{equation}
The estimator $D_L$ as defined in Eq.~(\ref{E:naive}) implies that:
\begin{equation} \label{E:DLDL'}
\langle D_L D_{L'} \rangle = \left( \frac{2 \pi}{A L L'} \right)^2
\frac{1}{L \Delta L L' \Delta L'} \int_{a_L} \frac{d^2 \vecla'}{(2 \pi)^2} \,
\int_{a_{L'}} \frac{d^2 \veclb'}{(2 \pi)^2} \, \langle \left[
\est(\vecla') \cdot \est(-\vecla') \right] \left[
\est(\veclb') \cdot \est(-\veclb') \right] \rangle \, ,
\end{equation}
which can be evaluated using the same integral over the eight-point
correlation function described in Eq.~(\ref{E:8pt}).  Whereas 60 of the 105
zeroth-order terms in $C_{L}^{\len\len}$ coming from this equation were
nonvanishing for the variance, for the covariance only 52 terms are nonzero
provided that the widths $\Delta L$ and $\Delta L'$ are chosen so that the
annuli $a_L$ and $a_{L'}$ do not overlap.  This leads to a result analogous
to Eq.~(\ref{E:zeroPV}),
\begin{eqnarray} \label{E:zeroCV}
&& \langle D_L D_{L'} \rangle = \frac{\noise(L)}{L^2} \frac{\noise(L')}{L'^2} +
\nonumber \\ && \quad \quad \quad \quad \quad \quad \quad
\frac{2}{A L^3 \Delta L L'^3 \Delta L'}
\int_{a_L} \frac{d^2 \vecla'}{(2 \pi)^2} \,
\int_{a_{L'}} \frac{d^2 \veclb'}{(2 \pi)^2} \, \frac{\norm^2(l_{1}^{\prime})
\norm^2(l_{2}^{\prime})}{(l_{1}^{\prime})^2 (l_{2}^{\prime})^2} \intk{1}
f_{\cmb\cmb}(\vecka, \veckb) P(\vecka, \veckb, \vecla', \veclb') \, .
\nonumber \\
\end{eqnarray}
Note that the eight terms missing from the covariance when compared to the
variance have altered the first term of Eq.~(\ref{E:zeroCV}), and that it was
precisely these terms that provided the dominant contribution to the
variance of Eq.~(\ref{E:FinVar}) when $\Delta L(L)$ was chosen appropriately.
We therefore find that to zeroth order in $C_{L}^{\len\len}$, the
covariance is given by
\begin{equation} \label{E:FinCov}
\sigma_{\Pest \Pestp} = \frac{2}{A L \Delta L L' \Delta L'}
\left( \frac{\noise(L)}{L^2} \frac{\noise(L')}{L'^2} \right)^2
\int_{a_L} \frac{d^2 \vecla'}{(2 \pi)^2} \,
\int_{a_{L'}} \frac{d^2 \veclb'}{(2 \pi)^2} \, \intk{1}
f_{\cmb\cmb}(\vecka, \veckb) P(\vecka, \veckb, \vecla', \veclb') \, ,
\end{equation}
where we have extracted the $\norm(l_{i}^{\prime})$ from the annular
integrals since they are slowly varying over the widths $\Delta L$ and
$\Delta L'$.  For the same reasons that terms of this form were a subdominant
contribution to the variance as discussed previously, we expect the
covariance to be suppressed as well.  If we define the ratio
\begin{equation} \label{E:Rcov}
R_{LL^{\prime}} \equiv \frac{\sigma_{\Pest \Pestp}}{\sqrt{\sigma_{\Pest}^2
\sigma_{\Pestp}^2}} \, ,
\end{equation}
we can quantify this suppression.  The triple integral of Eq.~(\ref{E:FinCov}),
appearing in the numerator of $R_{LL^{\prime}}$, involves integration over
annuli with radii $L$ and $L^{\prime}$ and one integration over all Fourier
space.  The triple integrals in the variances $\sigma_{\Pest}^2$ and
$\sigma_{\Pestp}^2$ appearing in the denominator of $R_{LL^{\prime}}$ each
consist of a single integration over an annulus of radius $L$ and $L^{\prime}$
respectively and two integrations over all Fourier space.  If we make
the crude assumption that the integrand is constant, the ratio
$R_{LL^{\prime}}$ will simply be the ratio of these areas,
\begin{equation} \label{E:Rcovest}
R_{LL^{\prime}} \simeq \sqrt{(2 \pi L \Delta L)(2 \pi L^{\prime} \Delta
L^{\prime})}/\pi l_{\rm{max}}^2.
\end{equation}

\begin{figure}[t]
%\centerline{\psfig{file=CLcovPlot.ps,width=7.0in,angle=0}}
\begin{center}
\epsfig{file=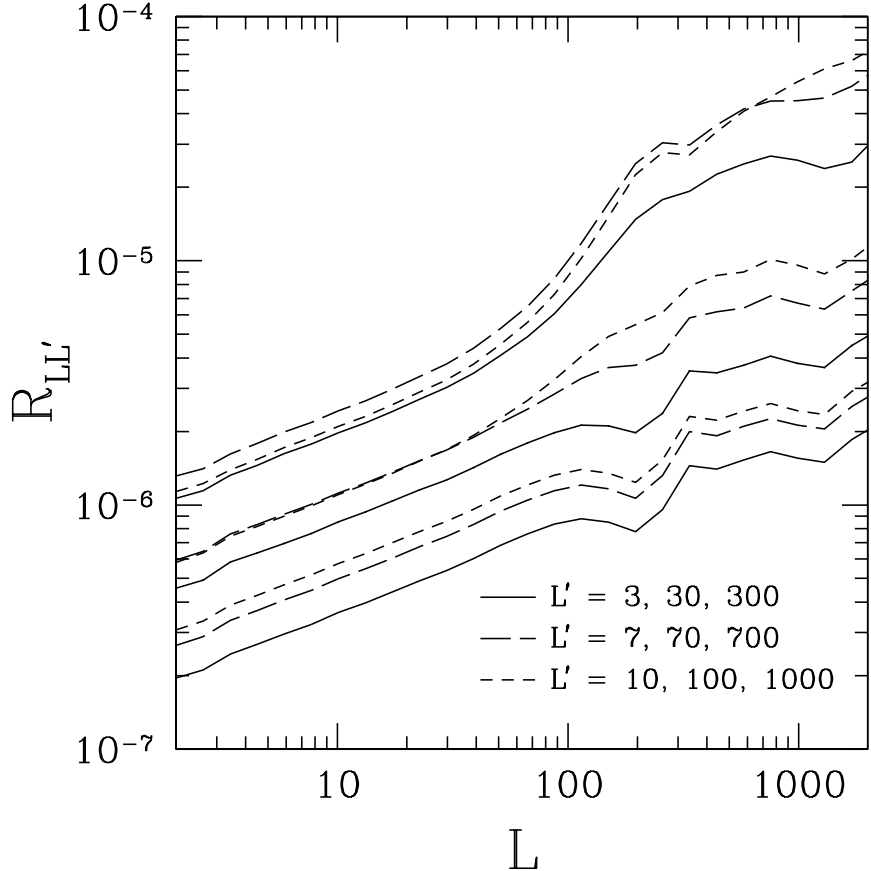,height=12cm,bbllx=18,bblly=144,bburx=570,bbury=715,clip=}
\caption{The ratio $R_{LL^{\prime}}$ for Planck as a function of $L$ for fixed
values of $L^{\prime}$.  The solid curves correspond to $L^{\prime} = 3, 30,
300$ ascending from bottom to top, while the long-dashed and short-dashed
curves correspond to $L^{\prime} = 7, 70, 700$ and $L^{\prime} = 10, 100, 1000$
respectively, again with curves in each sequence appearing from bottom to top
in the figure.
}
\label{F:cov}
\end{center}
\end{figure}

The ratio $R_{LL^{\prime}}$ is evaluated numerically for Planck in
Fig.~\ref{F:cov} as a function of $L$ for various fixed values of $L^{\prime}$.
The estimators $\Pest$ and $\Pestp$ were chosen such that
$\Delta L = \Delta L^{\prime} = 1$, while integrals over Fourier space were
cut off at $l_{\rm{max}} = 5000$.  Substituting these values into
Eq.~(\ref{E:Rcovest}), we expect
$R_{LL^{\prime}} \simeq 8.0 \times 10^{-8} \sqrt{L L^{\prime}}$.  This
crude estimate is surprisingly close to the numerically obtained results of
Fig.~\ref{F:cov}; in particular the slope of the curves is approximately 1/2
on this log-log plot.  Even at $L \simeq 1000$, $R_{LL^{\prime}} \lesssim
4.0 \times 10^{-4}$ suggesting that covariance in power-spectrum estimation
can safely be neglected for Planck.  The estimate of Eq.~(\ref{E:Rcovest})
implicitly depends on the experimental resolution $\theta$ because the
integrand appearing in expressions for the variance and covariance decreases
rapidly for $L \gtrsim l_{\rm{max}} \simeq \pi/\theta$.  For future
experiments with better resolution than Planck, $l_{\rm{max}}$ will be higher
implying by Eq.~(\ref{E:Rcovest}) that covariance will be even more
negligible.  

\section{Discussion}
\label{S:disc}

Weak gravitational lensing induces non-Gaussian correlations between modes of
the observed CMB temperature map as shown in Eq.~(\ref{E:EA}).  These
correlations, and assumptions about the Gaussian nature of the primordial
CMB, can be used to construct several temperature and polarization-based
estimators of the Fourier modes $\bfd(\bfL)$ of the deflection field.  This
procedure was outlined in Ref.~\cite{HuOka01}, however in calculating the noise
associated with this reconstruction, an assumption was made that the observed
temperature map was Gaussian.  In the presence of lensing this assumption is
invalid; when calculating the variance of quadratic estimators all
permutations of the observed trispectrum must be taken into account.  One such
permutation reflects the desired correlation making our estimator sensitive
to $\bfd(\bfL)$, but the remaining two permutations induce additional variance
proportional to the lensing-potential power spectrum $C_{L}^{\len\len}$.
While subdominant, this variance will become increasing significant for future
experiments as shown in the right-hand panel of Fig.~\ref{F:Plrec}.  Since the
power spectrum $C_{L}^{\len\len}$ is itself a measure of uncertainty in the
deflection field, this additional variance in lensing reconstruction acts as
a bias during power-spectrum estimation because there is no {\it a priori} way
to distinguish it from the intrinsic variance of the underlying distribution.
Our calculation of the dependence of this variance on $C_{L}^{\len\len}$ allows
it in principle to be subtracted iteratively, which will prevent a systematic
$5-10 \%$ overestimate of $C_{L}^{\len\len}$ at low $L$.

We close by considering several possible observational obstacles to the
scheme for lensing reconstruction and power-spectrum estimation presented
above.  One hindrance is secondary
contributions to the CMB such as the SZ and ISW effects.  These effects
increase the total temperature power spectrum appearing in the denominator of
the optimum filter $\filt(\vecla,\veclb)$ of Eq.~(\ref{E:filt}) as would
additional instrumental noise.  They also correlate with the large-scale
structure at low redshifts inducing further non-Gaussian couplings and
additional variance to lensing reconstruction.
Fortunately for our purposes the frequency dependence of the thermal SZ effect
differs from that of a blackbody. It can therefore be separated in
principle from the lensed primordial CMB by an experiment with several
frequency channels \cite{Cooetal00}.  The ISW effect cannot be removed in this
manner, but is too small to significantly inhibit lensing reconstruction.
Polarization-dependent secondary effects are expected to appear at higher
orders in the density contrast \cite{Hu00b}, and we therefore anticipate that
they will not make a contribution at the levels considered here.  A potentially
more serious problem is that of galactic foregrounds, which though uncorrelated
with the lensing signal may be substantial at certain frequencies.  Significant
polarization has also been observed in some of these sources \cite{galactic}.
We hope to understand and minimize the effects of galactic foregrounds
in future work, and to pursue further refinements of lensing reconstruction.

\acknowledgements

We thank Wayne Hu for useful discussions. This work was supported in part by
NASA NAG5-11985 and DoE DE-FG03-92-ER40701.  Kesden
acknowledges the support of an NSF Graduate Fellowship and AC
acknowledges support from the Sherman Fairchild Foundation.

\appendix
\section{Polarization-based Estimators}
\label{S:App}

Here we provide the appropriate formulas for deriving the variance associated
with polarization-based estimators of the deflection field $\bfd(\bfL)$.
The CMB polarization can be decomposed into E and B-modes
\cite{KamKosSte97}.  These modes are mixed by weak lensing
such that to linear order in $\len(\bfL)$,
\begin{eqnarray}
\tilde E(\vecl) &=& E(\vecl) - \intl{1} \left[E(\vecl_1)
\cos 2 (\varphi_{\vecl_1} -\varphi_\vecl) - B(\vecl_1)
\sin 2 (\varphi_{\vecl_1} -\varphi_\vecl)\right] \phi(\vecl-\vecl_1)
\left[ (\vecl-\vecl_1) \cdot \vecl_1 \right] \, ,\nonumber \\
\tilde B(\vecl) &=& B(\vecl) - \intl{1} \left[E(\vecl_1)
\sin 2 (\varphi_{\vecl_1} -\varphi_\vecl) + B(\vecl_1)
\cos 2 (\varphi_{\vecl_1} -\varphi_\vecl)\right] \phi(\vecl-\vecl_1)
\left[ (\vecl-\vecl_1) \cdot \vecl_1 \right] \, .
\end{eqnarray}
We can exploit the sensitivity of the polarization modes to the lensing
potential to construct lensing estimators from quadratic combinations of
polarization modes.  We generalize Eq.~(\ref{E:EA}) to arbitrary combinations
$\{X, X'\}$ of $\cmb$, E, and B-modes as first derived in
Ref.~\cite{HuOka01},
\begin{equation} \label{E:XX}
\langle X^\tot(\bfl) X'^\tot(\bflp) \rangle_{\rm{CMB}} =
f_{XX'}(\bfl, \bflp) \phi(\bfL) \, ,
\end{equation}
where
\begin{eqnarray} \label{E:XXbias}
f_{\cmb E}(\bfl, \bflp) &=& C_{l}^{\cmb E} \cos 2 (\varphi_{\bfl}
-\varphi_\bflp) (\bfL \cdot \bfl) +
C_{l^{\prime}}^{\cmb E} (\bfL \cdot \bflp) = f_{E \cmb}(\bflp, \bfl)
\, , \nonumber \\
f_{\cmb B}(\bfl, \bflp) &=& C_{l}^{\cmb E} \sin 2 (\varphi_{\bfl}
-\varphi_\bflp) (\bfL \cdot \bfl) = f_{B \cmb}(\bflp, \bfl)
\, , \nonumber \\
f_{EE}(\bfl, \bflp) &=& \left[ C_{l}^{EE} (\bfL \cdot \bfl) +
C_{l^{\prime}}^{EE} (\bfL \cdot \bflp) \right] 
\cos 2 (\varphi_{\bfl} -\varphi_\bflp) \, , \nonumber \\
f_{EB}(\bfl, \bflp) &=& \left[ C_{l}^{EE} (\bfL \cdot \bfl) +
C_{l^{\prime}}^{BB} (\bfL \cdot \bflp) \right] 
\sin 2 (\varphi_{\bfl} -\varphi_\bflp) = f_{BE}(\bflp, \bfl)
\, , \nonumber \\
f_{BB}(\bfl, \bflp) &=& \left[ C_{l}^{BB} (\bfL \cdot \bfl) +
C_{l^{\prime}}^{BB} (\bfL \cdot \bflp) \right] 
\cos 2 (\varphi_{\bfl} -\varphi_\bflp) \, . \nonumber \\
\end{eqnarray}
In deriving these results we used parity considerations to demand
$C_{l}^{\cmb B} = C_{l}^{EB} = 0$.
Using these relations, we follow the approach of Ref.~\cite{HuOka01}
to derive symmetric lensing estimators
\begin{equation} \label{E:estXX}
\estXX(\bfL) \equiv \frac{i\bfL \normXX(L)}{L^2} \intl{1} X^\tot(\vecla)
X^\tot(\veclb) \filtXX(\vecla, \veclb) \, ,
\end{equation}
and
\begin{equation} \label{E:estXX'}
\estXXp(\bfL) \equiv \frac{i\bfL \normXXp(L)}{L^2} \intl{1} \frac{1}{2} \left[
X^\tot(\vecla) X'^\tot(\veclb) + X'^\tot(\vecla) X^\tot(\veclb) \right]
\filtXXp(\vecla, \veclb) \, .
\end{equation}
We have explicitly symmetrized our estimators for $X \ne X'$ to simplify the
form of the optimal filter.  The normalization bias of the estimators is
removed by choosing
\begin{equation} \label{E:normXX}
\normXX(L) \equiv L^2 \left[ \intl{1} f_{XX}(\vecla, \veclb)
\filtXX(\vecla, \veclb) \right]^{-1} \, ,
\end{equation}
and
\begin{equation} \label{E:normXX'}
\normXXp(L) \equiv L^2 \left[ \intl{1} \frac{1}{2} \left[
f_{XX'}(\vecla, \veclb) + f_{XX'}(\veclb, \vecla) \right]
\filtXXp(\vecla, \veclb) \right]^{-1} \, .
\end{equation}
The minimum-variance filters $\filtXXp(\vecla, \veclb)$ for the various cases
$\{X, X'\}$ are given by
\begin{eqnarray} \label{E:filtXX'}
F_{\cmb E}(\vecla, \veclb) &=& \frac{f_{\cmb E}(\vecla, \veclb)
+ f_{\cmb E}(\veclb, \vecla)}
{C_{l_1}^{\cmb\cmb\tot} C_{l_2}^{EE\tot} + 2 C_{l_1}^{\cmb E \tot}
C_{l_2}^{\cmb E \tot} + C_{l_1}^{EE\tot} C_{l_2}^{\cmb\cmb\tot}}
\, , \nonumber \\
F_{\cmb B}(\vecla, \veclb) &=& \frac{f_{\cmb B}(\vecla, \veclb)
+ f_{\cmb B}(\veclb, \vecla)}
{C_{l_1}^{\cmb\cmb\tot} C_{l_2}^{EE\tot} + 
C_{l_1}^{EE\tot} C_{l_2}^{\cmb\cmb\tot}}
\, , \nonumber \\
F_{EE}(\vecla, \veclb) &=& \frac{f_{EE}(\vecla, \veclb)}
{2 C_{l_1}^{EE\tot} C_{l_2}^{EE\tot}} \, , \nonumber \\
F_{EB}(\vecla, \veclb) &=& \frac{f_{EB}(\vecla, \veclb)
+ f_{EB}(\veclb, \vecla)}
{C_{l_1}^{EE\tot} C_{l_2}^{BB\tot}
+ C_{l_1}^{BB\tot} C_{l_2}^{EE\tot}}
\, , \nonumber \\
F_{BB}(\vecla, \veclb) &=& \frac{f_{BB}(\vecla, \veclb)}
{2 C_{l_1}^{BB\tot} C_{l_2}^{BB\tot}} \, . \nonumber \\
\end{eqnarray}
Using these optimal filters for the estimators defined in Eqs.~(\ref{E:estXX})
and (\ref{E:estXX'}), we can calculate the variances for these estimators in a
fashion entirely analogoug to Eq.~(\ref{E:estvar}),
\begin{eqnarray} \label{E:XXvar}
&& \Big\langle \langle \estXX^{\ast}(\bfL) \cdot \estXX(\bfLp)
\rangle_{\rm{CMB}} - \langle \estXX^{\ast}(\bfL) \rangle_{\rm{CMB}} \cdot
\langle \estXX(\bfLp) \rangle_{\rm{CMB}} \Big\rangle_{\rm{LSS}}
= \frac{\normXX(L)}{L}
\frac{\normXX(L^{\prime})}{L^{\prime}} \nonumber \\
&& \quad \times \intl{1} \intl{1'} \langle
X^\tot(-\vecla) X^\tot(-\veclb) X^\tot(\vecla') X^\tot(\veclb')
\rangle \filtXX(\vecla, \veclb) \filtXX(\vecla', \veclb')
- (2 \pi)^2 \delta_\dirac(\bfL - \bfLp) C_{L}^{dd}
\, , \nonumber \\
\end{eqnarray}
\begin{eqnarray} \label{E:XX'var}
&& \Big\langle \langle \estXXp^{\ast}(\bfL) \cdot \estXXp(\bfLp)
\rangle_{\rm{CMB}} - \langle \estXXp^{\ast}(\bfL) \rangle_{\rm{CMB}} \cdot
\langle \estXXp(\bfLp) \rangle_{\rm{CMB}} \Big\rangle_{\rm{LSS}}
= \frac{\normXXp(L)}{L} \frac{\normXXp(L^{\prime})}{L^{\prime}} \intl{1}
\intl{1'} \nonumber \\
&& \quad \quad \times
\frac{1}{4} \langle \left[ X^\tot(-\vecla) X'^\tot(-\veclb) +
X'^\tot(-\vecla) X^\tot(-\veclb) \right] \left[ X^\tot(\vecla')
X'^\tot(\veclb') + X'^\tot(\vecla') X^\tot(\veclb') \right]
\rangle \filtXXp(\vecla, \veclb) \filtXXp(\vecla', \veclb') \nonumber \\
&& \quad \quad - (2 \pi)^2 \delta_\dirac(\bfL - \bfLp) C_{L}^{dd}
\, . \nonumber \\
\end{eqnarray}
As for that of the temperature estimator, these variance will consist of
zeroth-order terms in $C_{L}^{\len\len}$, $\noiseXX(L) = \normXX(L)$ and
$\noiseXXp(L) = \normXXp(L)$, and first-order terms,
\begin{eqnarray} \label{E:1stXX}
&& \noiseXXP(L) =  \frac{\normXX^2(L)}{L^2} \intl{1} \intl{1'}
\filtXX(\vecla, \veclb) \filtXX(\vecla', \veclb') \nonumber \\
&& \quad \quad \times \Big\{
C_{|\vecla - \vecla'|}^{\phi\phi} f_{XX}(-\vecla, \vecla')
f_{XX}(-\veclb, \veclb') +
C_{|\vecla-\veclb'|}^{\phi\phi} f_{XX}(-\vecla, \veclb')
f_{XX}(-\veclb, \vecla') \Big\} \, , \nonumber \\
\end{eqnarray}
\begin{eqnarray} \label{E:1stXX'}
&& \noiseXXpP(L) =  \frac{\normXXp^2(L)}{L^2} \intl{1} \intl{1'}
\filtXXp(\vecla, \veclb) \filtXXp(\vecla', \veclb') \nonumber \\
&& \quad \quad \times \frac{1}{4} \Big\{
C_{|\vecla - \vecla'|}^{\phi\phi}
\bigl[ f_{XX}(-\vecla, \vecla') f_{X'X'}(-\veclb, \veclb') +
f_{XX'}(-\vecla, \vecla') f_{X'X}(-\veclb, \veclb') \nonumber \\
&& \quad \quad \quad \quad
+ f_{X'X}(-\vecla, \vecla') f_{XX'}(-\veclb, \veclb') +
f_{X'X'}(-\vecla, \vecla') f_{XX}(-\veclb, \veclb') \bigl] \nonumber \\
&& \quad \quad \quad
+ C_{|\vecla-\veclb'|}^{\phi\phi} 
\bigl[ f_{XX'}(-\vecla, \veclb') f_{X'X}(-\veclb, \vecla') +
f_{XX}(-\vecla, \veclb') f_{X'X'}(-\veclb, \vecla') \nonumber \\
&& \quad \quad \quad \quad
+ f_{X'X'}(-\vecla, \veclb') f_{XX}(-\veclb, \vecla') +
f_{X'X}(-\vecla, \veclb') f_{XX'}(-\veclb, \vecla') \bigl]
\Big\} \, , \nonumber \\
\end{eqnarray}
The filters given in Eq.~(\ref{E:filtXX'}) are no longer
optimal in the presence of this additional noise, but the
difference between these filters and the optimal filters should be negligible
provided that $\noiseXXP(L) \ll \noiseXX(L)$, $\noiseXXpP(L) \ll \noiseXXp(L)$.
For the purposes of power-spectrum estimation, the terms $\noiseXXP(L)$ and
$\noiseXXpP(L)$ are not only an additional contribution to the variance, but
are also a systematic bias if not subtracted iteratively following
Eq.~(\ref{E:naive3}).

A final point to consider is that the six different estimators $\estXXp(\bfL)$
defined in this paper are not independent, as they are constructed from only
three distinct maps.  The covariance matrix for the six estimators will
therefore not be diagonal, and this needs to be taken into account if the
estimators are to be linearly combined to produce a single minimum-variance
estimator.  The off-diagonal elements of the covariance matrix can be
evaluated in a straightforward manner involving pairs of double integrals
similar to those of Eqs.~(\ref{E:XXvar}) and (\ref{E:XX'var}).

\end{document}